\documentclass[12pt]{book}

\usepackage[dvips]{graphicx,color}
\usepackage{makeidx,tsukuba}

\makeauthorindex
\makeindex

\begin{document}

\BookTitle{\itshape The 28th International Cosmic Ray Conference}
\CopyRight{\copyright 2003 by Universal Academy Press, Inc.}
\CopyRight{\copyright 2003 by Universal Academy Press, Inc.}
\pagenumbering{arabic}
\chapter{Search for Sub-TeV Gamma Rays Coincident with BATSE Gamma Ray
Bursts}
\author{John Poirier, Christopher D'Andrea, Joseph Gress, and Doran Race\\
{\it Center for Astrophysics at Notre Dame, University of Notre Dame,\\
Notre Dame, IN 46556, USA} }

\section*{Abstract}
Project {\small GRAND} is a 100m $\times$ 100m air shower array of
proportional
wire chambers (PWCs).
There are 64 stations each with eight 1.29~m$^2$ PWC planes
arranged in four orthogonal pairs placed vertically above one another to
geometrically measure the angles of charged secondaries.
A steel plate above the
bottom pair of PWCs differentiates muons (which pass undeflected through the
steel) from non-penetrating particles.  FLUKA Monte Carlo studies show that a
TeV
gamma ray striking the atmosphere at normal incidence produces 0.23 muons
which
reach ground level where their angles and identities are measured.  Thus,
paradoxically, secondary muons are used as a {\it signature} for gamma ray
primaries.  The data are examined for possible angular and time coincidences
with
eight gamma ray bursts (GRBs) detected by BATSE.  Seven of the GRBs were
selected
because of their good acceptance by {\small GRAND} and high BATSE fluence.
The eighth GRB was added
due to its possible coincident detection by Milagrito.
For each of the eight candidate GRBs, the
number of excess counts during the BATSE T90 time interval and within $\pm
5^{\circ}$ of BATSE's direction was obtained.
The highest statistical
significance reported in this paper ($2.7\sigma$) is for the event that was
predicted to be the most likely to be observed (GRB 971110).
 \section{Introduction}
The mystery of the astrophysical origin for gamma bursts (GRBs)
has been present for some time.  Although there is no one complete paradigm
for
their origin, it is likely that energetic ($\sim$ TeV) gamma rays
might also be produced along with the low energy burst.
Previous literature presents at least some evidence for a
possible association of energetic gamma rays with low-energy GRBs.  EGRET
detected several GRBs which emitted high energy photons in the $\sim$ 100 MeV
to
18 GeV range [2,3,12].  There have
also been some results from the Tibet air shower array suggestive of gamma
rays
beyond the TeV range [1,8].  Possible evidence for TeV emission in
coincidence
with a BATSE GRB has been reported from the Milagrito detector [5,6,7].

\section {Project {\small GRAND}}
{\small GRAND} is located at $86.2^{\circ}$ W and $41.7^{\circ}$ N.  It
detects
cosmic
ray
secondaries at ground level by means of 64 tracking stations of proportional
wire
chambers (PWCs).
Each station has eight 1.29 m$^2$ PWC planes.
Each secondary muon is measured to $0.26^{\circ}$ precision, on average, in
each of two orthogonal planes.
A 51 mm thick steel plate is located between the sixth and seventh PWC plane
allowing
muons to be distinguished from electrons.  The data presented here are from
single-track triggers with only the muon tracks selected.  Secondary muons
are
primarily the result of interactions of primary cosmic rays with the
atmosphere
producing pions, which then decay into muons.  The muons are produced at
small
angles relative to the pion and are then deflected in the earth's magnetic
field and scattered in the atmosphere resulting in an effective net angular
resolution of about $\pm 5^{\circ}$ (depending slightly on the primary
spectral
index; see Fig 6 in [10]).  Thus the experimental angular resolution is much
better than the effective angular resolution which is governed by the
deflections suffered by the muons before they reach the detector.

{\small GRAND} utilizes the fact that gamma rays have a detectable signal of
muons from
gamma-hadro production in the atmosphere making it possible to study
coincidences between GRBs and gamma ray showers in the 10 GeV $\le$
E$_\gamma$
$\le$ 1 TeV
energy region (see Fig 6 of [10]).

In the gamma primary energy region above
10~GeV, the recorded muon rate is 2400 Hz.
A FLUKA MC simulation shows that a 1 TeV gamma ray normally incident upon the
earth's
atmosphere produces an average of 0.23 muons which reach detection
level.  Paradoxically,
muons are used as a {\it signal} for gamma ray primaries.
Our response to primary gamma rays has
a threshold of $\sim$ 10~GeV (optimum response).
Our ability to correlate short bursts of muons with an identifiable source
of
primary radiation has been shown in a detection which was coincident with the
solar
flare of 15 April 2001 [9].  The statistical significance of this
observation was 6.1$\sigma$ for a ground level event of 0.6 hours duration.
 \section{Data Analysis}

For each GRB, an acceptance factor, based on the elevation angle of the burst
and
the geometry of the array, was multiplied by BATSE's fluence in their highest
energy bin to obtain a
rudimentary likelihood, $LogLk$, that we could observe the GRB.
The top seven GRBs for which we had data were selected for analyses.
In addition the  Milagrito event, GRB 970417a,  was included.

\begin{table}
\caption{Summary of the Eight Events Analyzed}
\begin{center}
\begin{tabular}{lcccccccccc}
\hline
\hline
GRB & Trig & T90 & RA & Dec & $\delta\theta$ & Elev& $LogLk$ &
$N_\mu\pm\sigma_{Tot}$ \\
\hline
971110  & 6472 & 195.2 & 242 & 50 & 0.6 & 81 & 5.18 & $466 \pm 171$ \\
990123  & 7343 &  62.5 & 229 & 42 & 0.4 & 56 & 5.13 & $3   \pm 36$  \\
940526  & 2994 &  48.6 & 132 & 34 & 1.7 & 66 & 4.68 & $20  \pm 28$  \\
980420  & 6694 &  39.9 & 293 & 27 & 0.6 & 68 & 4.02 & $39  \pm 47$  \\
960428  & 5450 & 172.2 & 304 & 35 & 1.0 & 70 & 3.83 & $57  \pm 78$  \\
980105  & 6560 &  36.8 & 37  & 52 & 1.4 & 79 & 3.46 & $-15 \pm 61$  \\
980301  & 6619 &  36.0 & 148 & 35 & 1.3 & 76 & 3.17 & $38  \pm 56$  \\
970417a & 6188 &  7.9  & 290 & 54 & 1.6 & 62 & 2.08 & $20  \pm 17$  \\
\hline
\hline
\end{tabular}
\end{center}
\end{table}

First, the BATSE RA and Dec
were transformed to a local horizon coordinate system and
projected onto the xz-plane ($\theta_x$) and the yz-plane ($\theta_y$).
A window of
$\pm5^o$ in $\Delta \theta_x$ and $\Delta \theta_y$ was centered on the
location of the
GRB.
To correct for the experimental dead time, the total event rate over the
whole sky was employed as a high-statistics measure of the live time of each
time bin;
each bin's data were corrected for its corresponding live time.
The background was determined  during a
time interval of 20$\times$T90 before the start of the BATSE trigger (except
for GRB 971110 which was 10$\times$T90 in order to stay within the
data tape for this event).

Table 1 lists BATSE data on eight gamma ray bursts: the date of the
trigger, the trigger
number, the time duration for 90\% of the burst's counts to occur (in
seconds); the right ascension, declination, and the BATSE angular error in
degrees.  Next is the angular elevation (degrees) above our horizon and our
selection
criterion ($LogLk$), described below.  The last column of Table 1 summarize
the
muon secondaries observed by us within a $\pm 5^{\circ}$ square angular
window centered on the burst location during the BATSE T90 time interval.
The
error is the total error (statistical plus systematic) which is discussed
below.
Further details on these observations can be found in [11].

\section{Significance and Errors}
The signal (Sig) calculated for a GRB is the difference between the total
counts inside the T90 interval (corrected for dead time) and the background
 counts normalized to the live time of the T90 time interval and corrected
for dead time.
The statistical significance (number of standard
deviations above background) of each event was determined according to the
likelihood
ratio method of Li and Ma, [4]
which makes a good
accounting of the true significance for events with differing background and
event times.

It is possible that the background fluctuates
in excess of the expected  $\sqrt{N}$ statistics.
As a check on systematic errors in the signal for GRB 971110, similar angular
sections
of the sky (which have the same absolute values of $\theta_x$ and $\theta_y$
and thus the same average counting rates) and the same time interval (T90)
but
at
different, neighboring times were analyzed.  These time intervals span
$\pm$26~hours
relative to the BATSE trigger for GRB 971110 and were analyzed
in the same way as described in the preceding paragraphs.
The standard
deviation width of the distribution of these 1587 background cases is 171,
whereas the expected statistical deviation
based upon the square root of the background count rate is only 141.
Adopting this  total error as the
quadrature sum of the statistical and systematic errors, then 97 counts are
ascribed
as the additional systematic error in our analysis for GRB 971110.
With this additional systematic error, the ratio of signal-to-noise becomes
$Sig/\delta Sig = 2.72$.
The fractional additional systematic error for GRB 971110 was then used for
the
other GRB candidates in order to estimate their total errors.
The final results with total errors (statistical plus systematic)
are in the last column of the Table 1.

The probability of a $+2.7\sigma$ fluctuation in an assumed Gaussian
white-noise
distribution
is $3.5 \times 10^{-3}$.
The probability without assuming a
Gaussian shape was measured with the 1587 background cases:
There were ten random
fluctuations $\ge |466|$ yielding
a probability of $3.2\times 10^{-3}$ for a background
fluctuation to produce a fluctuation of $\ge +466$.
For one event out of the  eight analyzed to exhibit this much deviation would
correspond to a
random probability of 0.025.
Thus, the statistics of this event are interesting but not compelling.

No convincing evidence is found, though there is a
possible 2.7$\sigma$ detection associated with the most probable candidate.

The authors wish to acknowledge contributions of
D. Baker, J. Carpenter, S. Desch, M. Dunford, P.C. Fragile, C.F. Lin, M.
L\'opez del Puerto,
G.J. Mathews, R. Skibba, J. Vermedahl, and M. Wysocki;
and satellite data from BATSE (NASA).
Project {\small GRAND}'s research is presently being funded through
the University of Notre Dame and private grants.

\section{References}
\vspace{\baselineskip}
\re
1.\ Amenomori \ et al. \ 1996, Astronomy and Astrophysics 311, 919
\re
2.\ Catelli J.R., Dingus, B.L., Schneid, E.J. \ 1997, AIP Conference Proc.
428
\re
3.\ Hurley K. \ 1994, Nature 372, 652
\re
4.\ Li T.P., Ma Y. Q. \ 1983, Astrophysical Journal 272, 317
\re
5.\ Milagrito Collab, Atkins R. \ et al.\ 2000, ApJ Lett. 553, L119
\re
6.\ Milagrito Collab, Atkins R. \ et al.\ 2003, ApJ. 583, 824
\re
7.\ Milagrito Collab, Leonor I.R. \ et al.\ 1999, Proc. of the 26th ICRC 4,
12
\re
8.\ Padilla L. \ et al. \ 1998, Astronomy and Astrophysics 337, 43
\re
9.\ Poirier J., D'Andrea C. \ 2002, Journal of Geophysical Research 107, 1376
\re
10.\ Poirier J., Roesler S., Fass\'o A., \ 2002, \ Astroparticle Physics 17,
441
\re
11.\ Poirier J. \ et al. \ 2002, Physical Review D 67, 042001
\re
12.\ Schneid E.J. \ et al. \ 1992, Astronomy and Astrophysics 255, L13
\endofpaper
\end{document}